\documentclass{IEEEtran}

\usepackage[table]{xcolor}
\usepackage{cite}
\usepackage{graphicx}
\usepackage{lipsum}
\usepackage{stfloats}
\usepackage{amssymb}
\usepackage{pifont}
\usepackage{bbding}
\usepackage{booktabs}
\usepackage{ctable}
\usepackage{threeparttable}
\usepackage{footmisc}
\usepackage{framed} 
\usepackage{array}
\usepackage{multirow}
\usepackage{longtable}
\usepackage{rotating}
\usepackage{fancyhdr}
\usepackage{lastpage}
\usepackage{layout}
\usepackage{wrapfig}
\usepackage{xcolor}
\usepackage{tabularx}
\usepackage{tabu}
\usepackage{enumitem}
\usepackage{url}
\graphicspath{{Figures/}}
\DeclareGraphicsExtensions{.png, .eps}

\begin{document}
	%\title{Integrating AI and Blockchain for Intelligent Safe-Driving Assessment in CAV Networks: A Smart and Secure Approach}
	\title{Smart and Secure CAV Networks Empowered by AI-Enabled Blockchain: The Next Frontier for Intelligent Safe Driving Assessment}
	\author{Le Xia, Yao Sun, Rafiq Swash, Lina Mohjazi, Lei Zhang, and Muhammad Ali Imran
	\thanks{
	
	Le Xia, Yao Sun (\textit{corresponding author}), Lina Mohjazi, Lei Zhang, and Muhammad Ali Imran are with University of Glasgow, United Kingdom;
	
	Rafiq Swash is with AIDrivers Ltd. and Brunel University London, United Kingdom.
	}}	
	\maketitle
	\begin{abstract}

	Securing safe driving for connected and autonomous vehicles (CAVs) continues to be a widespread concern, despite various sophisticated functions delivered by artificial intelligence for in-vehicle devices. Diverse malicious network attacks are ubiquitous, along with the worldwide implementation of the Internet of Vehicles, which exposes a range of reliability and privacy threats for managing data in CAV networks. Combined with the fact that the capability of existing CAVs in handling intensive computation tasks is limited, this implies a need for designing an efficient assessment system to guarantee autonomous driving safety without compromising data security. In this article we propose a novel framework, namely Blockchain-enabled intElligent Safe-driving assessmenT (BEST), which offers a smart and reliable approach for conducting safe driving supervision while protecting vehicular information. Specifically, a promising solution that exploits a long short-term memory model is introduced to assess the safety level of the moving CAVs. Then we investigate how a distributed blockchain obtains adequate trustworthiness and robustness for CAV data by adopting a byzantine fault tolerance-based delegated proof-of-stake consensus mechanism. Simulation results demonstrate that our presented BEST gains better data credibility with a higher prediction accuracy for vehicular safety assessment when compared with existing schemes. Finally, we discuss several open challenges that need to be addressed in future CAV networks.
	\end{abstract}
	
	\section{Introduction}

	With the proliferation of information demands among connected vehicles, maintaining wireless connectivity between vehicular networks and roadside infrastructures is becoming increasingly indispensable. In this setup, the connection is primarily carried out through specialized communication technologies. For example, roadside unit-based (RSU) dedicated short-range communications (DSRC) or base station-enabled cellular networks~\cite{abboud2016interworking}. Both safe road surveillance and reliable vehicle control can be further ensured by allowing vehicle-to-everything communication. Additionally, the advancement of artificial intelligence (AI) has garnered a great deal of significance in vehicular networks -- connected and autonomous vehicles (CAVs) -- to liberate humans physically and mentally from daily driving tasks. Thanks to intelligent navigation, automated scheduling, and orderly driving, the promotion of CAV applications not only mitigates traffic congestion and resource consumption, but also enforces travel effectiveness and even reduces the casualty rates of traffic accidents~\cite{de2020driverless,wang2018networking}. Nevertheless, the current CAV network still faces several challenges, which can be briefly attributed to the following two aspects.
	\begin{itemize}[leftmargin=*]
    \item \emph{Driving Safety}:
%    Since the ultimate goal of autonomous driving is to reach the fifth level, i.e., full automation as defined by the Society of Automotive Engineers~\cite{sae2014taxonomy}, the autonomy of the vehicle itself should be the most critical factor for safety.
%    %Through the above researches, we realize that there are still some challenges on the road to achieve the ultimate goal of autonomous driving, i.e. full automation defined by Society of Automotive Engineers~\cite{sae2014taxonomy}, which indicates the autonomy of the vehicle itself should be the most critical factor for safety.
%    However, a malfunction resulting from unexpected erroneous bugs or security breaches may cause catastrophic consequences, like severe safety incidents or even casualties, such as the Uber accident occurred in 2018~\cite{efrati2018uber}. 
	Since the ultimate goal of autonomous driving is to reach the fifth level (full automation) as defined by the Society of Automotive Engineers~\cite{sae2014taxonomy}, the autonomy of the vehicle itself should be the most critical factor for safety. However, a malfunction resulting from unexpected bugs or security breaches may cause catastrophic consequences, including severe safety incidents or even casualties, such as the Uber accident that occurred in 2018~\cite{efrati2018uber}.
    \item \emph{Data Security}:
%    The security and authenticity of vehicular data are also crucial for driving safety.
%    Unfortunately, current identification, authentication and management for vehicular information are all handled by third parties.
%    As a result of this centralized management architecture, the fears of data tampering and privacy leakage are growing notably and leading to a core problem in trust.
%% growing trust fears of data tampering and privacy leakage thus disseminate due to this centralized management architecture.
%    Furthermore, diverse malicious attacks on CAV networks also become pervasive nowadays (e.g., camera blinding and GPS jamming~\cite{xing2019trust}) with its unceasing application scale. 
	The security and authenticity of vehicular data are also crucial for driving safety. Unfortunately, current identification, authentication, and management for vehicular information are handled by third parties. As a result of this centralized management architecture, the fears of data tampering and privacy leakage are growing and leading to a core problem in trust. Further more, diverse malicious attacks on CAV networks have become pervasive (e.g., camera blinding and GPS jamming~\cite{xing2019trust}) with its unceasing application scale.
    \end{itemize}
    
%    In response to the aforementioned issues, the fusion of deep learning (DL) and blockchain techniques seems to be a promising solution here.
%    First, DL should be a necessity to solve complicated prediction problems with its powerful neural networks~\cite{ye2021joint}.
%    This can be applied as an attractive method to accurately supervise the driving status of CAVs and then exploit the obtained feedback to implement proper countermeasures to the misbehaving vehicles, thereby, efficiently preventing accidents.
%    %If driving status of CAVs can be accurately supervised by a smart assessment system, then implementing proper countermeasures to those misbehaving vehicles according to the feedback, which can be believed as an efficient way to prevent those accidents from happening.
%    Meanwhile, blockchain, as an authority-decentralized technique, leverages a distributed digital ledger that records authorized transactions in blocks without the need for a central trusted medium, which guarantees ample trustworthiness and credibility for vehicular data management~\cite{posner2021federated}.

	In response to the aforementioned issues, the fusion of deep learning (DL) and blockchain techniques appears to be a promising solution. DL, with its powerful neural networks~\cite{ye2021joint}, should be a necessity to solve complicated prediction problems. This can be applied as an attractive method to accurately supervise the driving status of CAVs and then exploit the obtained feedback to implement proper countermeasures to the misbehaving vehicles, thereby efficiently preventing accidents. Meanwhile, blockchain, as an authority decentralized technique, leverages a distributed digital ledger that records authorized transactions in blocks without the need for a central trusted medium, which guarantees ample trustworthiness and credibility for vehicular data management~\cite{posner2021federated}.
    
%    Nevertheless, a perfect rationale of how to integrate DL with blockchain is very critical for the CAV network design.
%    Considering the complex vehicular environments, the intrinsic feature that combines the two is the status information of mobile CAVs.
%    Therefore, we specially consider to take the status data as the core hub between DL and blockchain models.
%    On the one hand, these data are obviously indispensable for DL to accurately assess the vehicular safety level.
%    On the other hand, blockchain can authenticate these data to prevent fake or dishonest content from mixing into the network to cause chaos, as well as to provide stable data source for DL prediction with sufficient reliability.
%    To the best of our knowledge, no article has conducted the same research before.
%    In this context, an efficient network architecture integrated with DL and blockchain becomes meaningful for provisioning driving safety and data security, simultaneously. 
	Nevertheless, a perfect rationale of how to integrate DL with blockchain is critical for the CAV network design. Considering the complex vehicular environments, the intrinsic feature that combines the two is the status information of mobile CAVs. Therefore, we specially consider to take the status data as the core hub between DL and blockchain models. On the one hand, the data is obviously indispensable for DL to accurately assess the vehicular safety level. On the other hand, blockchain can authenticate data to prevent fake or dishonest content from mixing into the network to cause chaos, as well as provide a stable data source for DL prediction with sufficient reliability. To the best of our knowledge, no article has conducted the same research before. In this context, an efficient network architecture integrated with DL and blockchain becomes meaningful for provisioning driving safety and data security simultaneously.
    
    %Correspondingly, an urgent need of integrating DL with blockchain into the CAV networks arises for simultaneously assuring a sufficiently driving-safe and data-secure circumstance.
    
    \begin{figure*}[ht]
		\centering
		\includegraphics[height=0.5150\textwidth, width=1\textwidth]{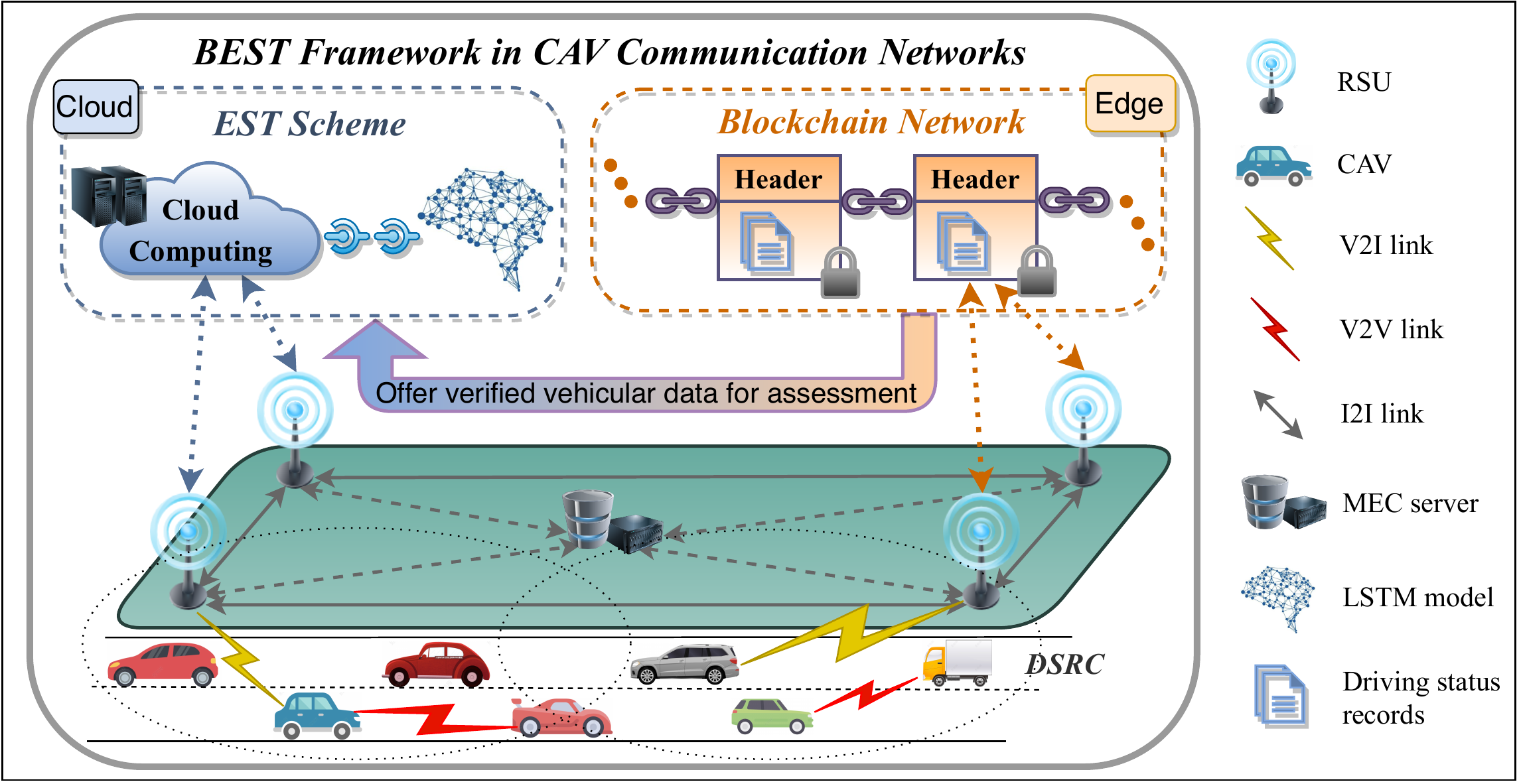} 
		\caption{An overview of the integrated BEST framework for CAV networks.}
		\label{Overview}
    \end{figure*}

	In this article, we propose a novel framework of Blockchain-enabled intElligent Safe-driving assessmenT (BEST) for CAV networks, as shown in Fig.~\ref{Overview}. Specifically, BEST comprises two components: an intElligent Safe-driving assessmenT (EST) scheme and a blockchain network underpinning a data management platform. For the EST scheme, a long short-term memory (LSTM) model is first adopted to cope with time series related prediction problems and align with high vehicular dynamics. By analyzing the driving status in different time slots, each CAV can receive a current safety level from the LSTM, as well as potential countermeasures, which can be processed and executed in the cloud. In addition, we further introduce a consortium blockchain to guarantee information security and privacy, which is supported by a mobile edge computing (MEC) technique to alleviate the computational pressure. With the implementation of byzantine fault tolerance-based delegated proof-of-stake (BFT-DPoS) consensus mechanism, we store real-time driving status data in multiple blocks at a fast block generation speed, which not only makes vehicular information immutable and unforgeable, but also serves the EST scheme with data authentication and traceability. Moreover, simulations are conducted to compare the performance of BEST with existing schemes. The results show that our BEST can effectively avoid the false information sharing from malicious CAVs, and simultaneously assess the safety level for each CAV with a high accuracy. Finally, we outline several challenges and prospects of BEST from the perspectives of incentive, efficiency, and resource utilization in CAV networks.
	
	For the remainder of this article, we first give an overview of the conventional CAV network along with several current obstacles to emphasize the significance of our BEST framework. Then, we specifically demonstrate how the LSTM performs safety level assessment to cooperate with the blockchain system. Afterward, the BEST is verified and discussed by simulation results. Finally, we open the doors for future direction and close with conclusions.
    
%    For the remainder of this article, we first give an overview of the conventional CAV network along with several current obstacles to emphasize the significance of our BEST framework.
%    Then, we specifically demonstrate how the LSTM performs safety level assessment to cooperate with the blockchain system. 
%    Afterward, the BEST is verified and discussed by simulation results.
%    Finally, we open the doors for future directions and close this article with conclusions.
	
	\section{Overview of CAV Networks and BEST Framework}
%	In this section, the fundamental architecture and existing limitations of current CAV networks are presented, followed by the overview descriptions of the BEST framework. 
	\subsection{Connected and Autonomous Vehicular Networks}
%	As the rapid development of AI and communication technology, it is foreseeable that the application scale of vehicular networks will rapidly expand in the upcoming years.
	First, we briefly introduce some core elements to provide a deeper insight of CAV networks, which are listed as follows.
	
	\begin{itemize}[leftmargin=*]
    \item \emph{CAVs:}
%    Undoubtedly, CAVs excel due to numerous advanced applications supported by AI.
%	The merits of AI are leveraged in numerous applications supported by CAVs.
%    By analyzing information gathered from multiple in-vehicle devices, vehicles can map out the optimal driving trajectory followed by intelligent decision execution, including tire orientation control and the change of lane or velocity.
	The merits of AI are leveraged in numerous applications supported by CAVs. By analyzing information gathered from multiple in-vehicle devices, vehicles can map out the optimal driving trajectory followed by intelligent decision execution, including tire orientation control and the change of lane or velocity.
    
    \item \emph{RSUs:}
%    The RSU refers to a core roadside infrastructure that performs data access functions for CAVs within its signal coverage, and also offers bi-directional communication for vehicles and other associated servers.
    The RSU refers to a core roadside i structure that performs data access functions for CAVs within its signal coverage, and also offers bi-directional communication for vehicles and other associated servers.
    
    \item \emph{Communication Networks:}
%    Communication is an indispensable technical element.
%    Sophisticated inter-vehicle communications provide multiple feasible options in automotive networking community.
%    For instance, a CAV uses its on-board units to wirelessly connect with other CAVs via vehicle-to-vehicle links, or to access an adjacent RSU using DSRC characterized by short-distance and low-latency communications. 
%    Similarly, I2I communications are also essential for ensuring a seamless connection between infrastructures, like RSUs, for better information interactions and a broader network coverage.
%    It is such compatibility and connectivity of AI and communication technologies that make vehicles smarter with unprecedented success.
	Sophisticated inter-vehicle communications provide multiple feasible options in the automotive networking community. For instance, a CAV uses its onboard units to wirelessly connect with other CAVs via vehicle-to-vehicle links, or to access an adjacent RSU using DSRC characterized by short-distance and low-latency communications.
    \end{itemize}
    
   % Despite its bright future, a number of risks are still hidden in the dark corner. 
    %On the one hand, autonomous driving mainly relies on sensors and networks control, which indicates the unknown malfunctions may occur at any time~\cite{cui2019review}. 
    %On the other hand, although RSUs build a feasible bridge between CAVs and the Internet, it also brings the possibilities for those malicious attackers at the same time.
    
%    Despite its bright future, a number of risks are still hidden in the dark corner. 
%	Despite its promising prospects, there still exists several critical challenges in the corner.
%    Generally, autonomous driving mainly relies on sensors and networks control, making CAVs susceptible to unknown malfunctions at any time~\cite{cui2019review}. 
%    Besides, although RSUs build a feasible bridge between CAVs and the Internet, they expose the CAV network to possible malicious attacks.
%    Here, we list some existing challenges in the CAV network below for summary. 
	Despite its promising prospects, there still exists several critical challenges. Generally, autonomous driving mainly relies on sensors and network controls, making CAVs susceptible to unknown malfunctions at any time~\cite{cui2019review}. Although RSUs build a feasible bridge between CAVs and the Internet, they expose the CAV network to possible malicious attacks. Here, we list some existing challenges in the CAV network for summary.
    
    \textbf{Limitations in Vehicular Capability:} 
%    Since the burden of data gathering and processing has been drastically increasing, it sets high standards in data traffic tradeoff to car manufacturers and brings colossal pressure for CAV design. 
%    Obviously, there is still a long way for a single CAV to solely finalize such considerable computational tasks, owing to the dual limitations of costs and technical level of vehicles. 
	Due to the increasing burden of data generation and limited processing capability, it becomes challenging for a single CAV to simultaneously perform tremendous computation and communication tasks.
	Additionally, the inaccuracy and inefficiency of in-vehicle devices detection may lead to erroneous decision making, thereby endangering the safety of passengers or passersby in close proximity~\cite{jie2020game}.
%    Hence, implementing an external assessment system is crucial for supervising vehicular behaviors and applying corresponding measures in the real environment. 
    
    \textbf{Threats on Data Security and Privacy:}
%    In addition to the loopholes of CAV itself, it is more likely to suffer attacks from external networks, compared with manually driven vehicles. 
%    Note that the malicious attacks here can be considered as harmful network viruses or massive fraudulent data spread by individuals or organizations, thus to interfere with the normal operation of the network and achieve their illegal purposes, such as stealing private information or even forcibly seizing control of targeted vehicles.
%    Apart from this, attacks from malicious participants are ubiquitous as well, pretending to be normal vehicles or servers to sneak into the network and gaining benefits.
%    These diverse attacks render conventional data protection methods (e.g., cryptography) to be inefficient and inappropriate when applied to CAV networks.
%    Consequently, comprehensive considerations must be taken into account to effectively secure vehicular data.
	In addition to the loopholes of CAV itself, it is more likely to suffer attacks from external networks compared with manually driven vehicles. Note that malicious attacks can be considered as harmful network viruses or massive fraudulent data spread by individuals or organizations to interfere with the normal operation of the network and achieve illegal purposes, such as stealing private information or forcibly seizing control of targeted vehicles.
	Apart from this, attacks from malicious participants are ubiquitous, pretending to be normal vehicles or servers that sneak into the network and gain benefits. These diverse attacks render conventional data protection methods (e.g., cryptography) to be inefficient and inappropriate when applied to CAV networks.
    
    \textbf{Centralization of Network Management:}
%    Generally, CAV networks are maintained via third entities with opening access, which may incur inevitable trust and security fears for clients as a result of centralization that makes networks more vulnerable to single point of attacks from outside.
%%    Specifically, potential treacherous employees may deliberately disclose privacy or tamper data record, thus breeding such concerns.
%    Hypothetically, once the central sever is centrally damaged through the external attacks (e.g., distributed denial of service attack~\cite{zargar2013survey}), it may result in severe consequences like transportation system paralysis or immense economic losses. 
	Generally, CAV networks are maintained via third entities with opening access, which may incur inevitable trust and security fears for clients as a result of centralization that makes networks more vulnerable to single point of attacks from the outside.
	Hypothetically, once the central sever is centrally damaged through the external attacks (e.g., distributed denial of service attack~\cite{zargar2013survey}), it may result in severe consequences like transportation system paralysis or vast economic losses.
    Furthermore, as CAV networks continue to scale up, the centralized approach will become increasingly overwhelmed by handling and storing such massive data.
   
    \subsection{BEST Framework}
\begin{figure*}[ht]
		\centering
		\includegraphics[width=1\textwidth]{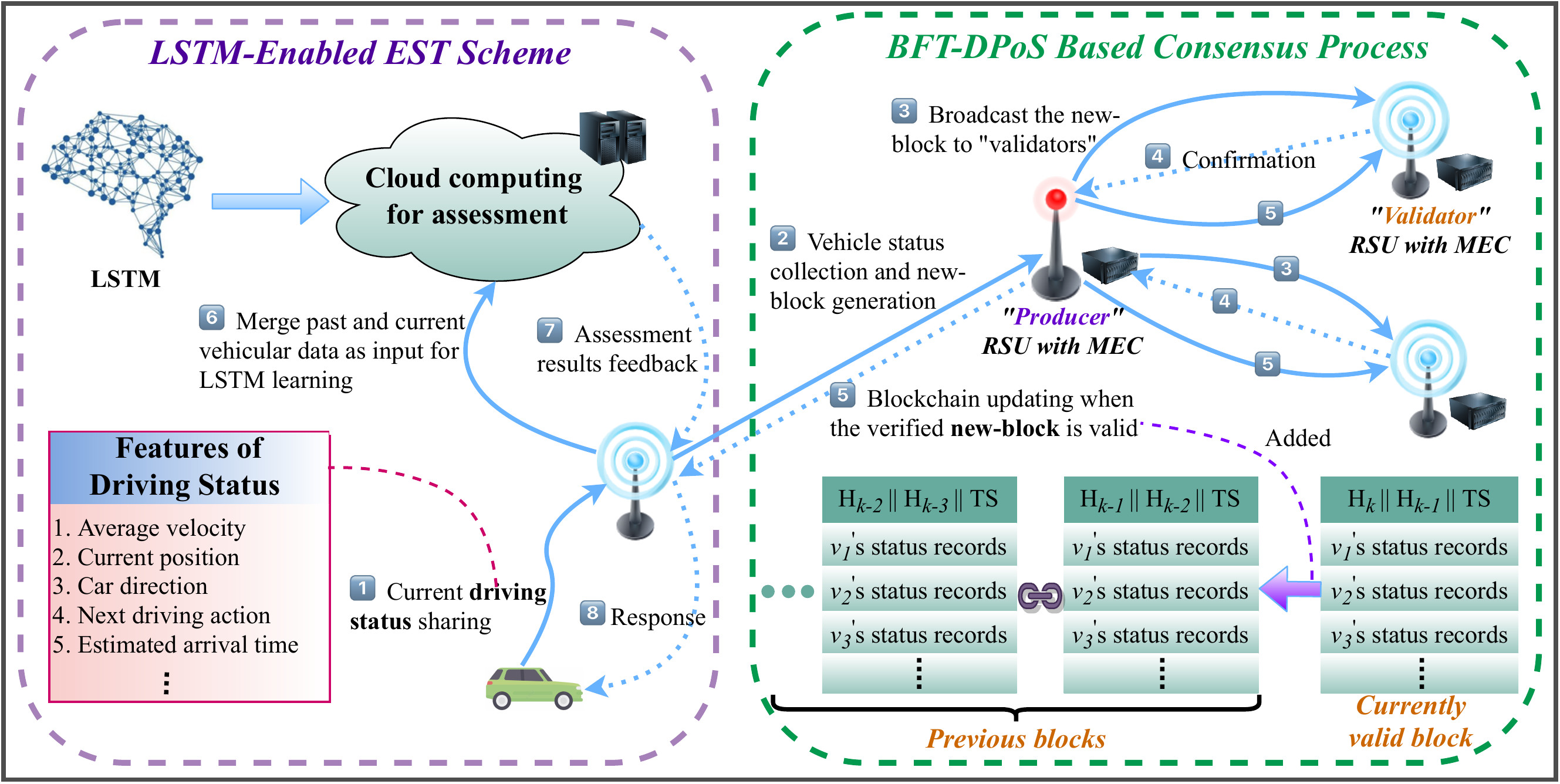} 
		\caption{Details of different phases of the proposed BEST framework in CAV networks.}
		\label{Architecture}
    \end{figure*}

    In order to tackle the challenging issues above, we propose a potential solution, namely the BEST framework for CAV networks.
    Notably, both components in BEST (EST scheme and blockchain network) are maintained and connected via RSUs.
    For the EST, an LSTM model is leveraged to assess the driving safety level of mobile CAVs, whereas each RSU is responsible for periodically collecting the driving status data of its covered vehicles.
%    Meanwhile, RSUs can also promptly apply countermeasures to the vehicles with misbehaviors, by either warning or performing artificial suspension.
%%	Here, we choose the LSTM algorithm to specially target the time series-related data in dynamic CAV scenarios.
%    Further, we integrate a scalable blockchain with the EST, in which transactions are securely encrypted, and the power that originally held in a third entity can be evenly decentralized to all RSUs.
%    Here, ``Transactions" can be interpreted as any information interaction in crypto between peers, mainly composed of the status records shared from mobile CAVs to RSUs in our roadmap, as depicted in Fig.~\ref{Architecture}.
%%    , the contents in $k_{th}$ block comprise the current block-hash $H_{k}$, with previous block-hash $H_{k-1}$, time stamp $TS$, and all data records occurred in a set of CAVs ($V=\left\{ v_{1},v_{2},\cdots \right\}$) within a given timeframe $T$.
%    The Hash encryption algorithm can guarantee the blockchain to defend against most of malicious attacks, making the attackers almost impossible to forge or alter ledger without being detected.
	Meanwhile, RSUs can also promptly apply countermeasures to the vehicles with misbehaviors, by either warning or performing artificial suspension. Further, we integrate a scalable blockchain with the EST, in which transactions are securely encrypted, and the power that was originally held in a third entity can be evenly decentralized to all RSUs.
	Here, ``transaction'' can be interpreted as any information interaction in crypto between peers, mainly composed of the status records shared from mobile CAVs to RSUs in our roadmap, as depicted in Fig.~\ref{Architecture}.
	The Hash encryption algorithm can guarantee the blockchain to defend against most malicious attacks, making the attackers almost impossible to forge or alter ledger without being detected.
	Note that each RSU participates in blockchain as a role of a blockchain node.
    Specially, all RSUs are functionally divided into two groups -- consensus RSU nodes (CRNs) and ordinary RSU nodes (ORNs) -- which are to match the adopted BFT-DPoS consensus mechanism.
    
%    The main focus of this article is to integrate them into one network to effectively address driving safety and data security issues at the same time.
%    Precisely, we believe that blockchain and AI can well complement each other in the BEST.
%	First, since the recorded status data are well-reserved and easily-traceable in chained blocks, this makes the blockchain a primary and reliable dataset for the LSTM.
%	Besides, the predicted results of LSTM on the vehicular safety level can also be stored in the blockchain, offering key information for the next round of prediction.
%    It is this very compatible internal collaboration between blockchain and DL that makes the entire CAV network more reliable and resilient.

	The main focus of this article is to integrate them into one network to effectively address driving safety and data security issues at the same time. Precisely, we believe that blockchain and AI can complement each other in the BEST. First, since the recorded status data are well-reserved and easily traceable in chained blocks, this makes the blockchain a primary and reliable dataset for the LSTM. Besides, the predicted results of LSTM on the vehicular safety level can also be stored in the blockchain, offering key information for the next round of prediction. It is this compatible internal collaboration between blockchain and DL that makes the entire CAV network more reliable and resilient.
    
%    Herein, we take the workflow of a CAV registered in the BEST as an exemplification to facilitate understanding. 
%    The vehicle is first required to accurately capture various status information through its multiple sophisticated sensors while driving. 
%	Next, the data are uploaded to the adjacent RSU, and the blockchain can thus verify them with the help of an effective consensus mechanism.
%	Afterwards, not only the current authenticated data, but also the data stored in blocks at the past time, will be input together into a well-designed and mature-trained LSTM model for assessing the specific level of driving safety and making countermeasures.
%	Furthermore, most computing tasks in blockchain process can be offloaded by deploying MEC on the clusters of RSUs for promoting system efficiency. 
%	Considering cost issues and relatively low-latency requirement of EST, the LSTM operation can be placed on remote cloud servers to firm ample computing resources while relieving the MEC servers and RSUs from extra computation load.
	
	Herein, we take the workflow of a CAV registered in the BEST as an exemplification to facilitate understanding. The vehicle is first required to accurately capture various status information through its multiple sophisticated sensors while driving. Next, the data are uploaded to the adjacent RSU, and the blockchain can verify the data with the help of an effective consensus mechanism. Afterward, the current authenticated data and the data stored in blocks at the past time will be input into a well-designed and fully trained LSTM model for assessing the specific level of driving safety and making countermeasures. Furthermore, most computing tasks in the blockchain process can be offloaded by deploying MEC on the clusters of RSUs for promoting system efficiency. Considering cost issues and the relatively low-latency requirement of EST, the LSTM operation can be placed on remote cloud servers to firm ample computing resources while relieving the MEC servers and RSUs from extra computation load.
	
	\section{Operations of BEST in CAV Networks}
%	In this section, we first illustrate the proposed EST scheme with an LSTM model to offer a smart and safe self-driving scenario. 
%	Then, the details of a blockchain system with its BFT-DPoS consensus mechanism applicable for CAV networks are presented, where its potential to assure data security and resilience for data management is highlighted.
	
	In this section, we first illustrate the proposed EST scheme with an LSTM model to offer a smart and safe self-driving scenario. Then, the details of a blockchain system with its BFT-DPoS consensus mechanism applicable for CAV networks are presented, where its potential to assure data security and resilience for data management is highlighted.
		
	\subsection{LSTM-Enabled EST Scheme}
	%Initially, AI is employed for trying to address complicated mathematical and computing problems. 
	%But as the development of AI has achieved unprecedented success, it also comes to contribute loads of efforts to the evolution of IoV. 
%	Powerful DL algorithms can be leveraged to perform intelligent applications, ranging from object detection and trajectory scheduling to video surveillance, to overcome the burdensome vehicular tasks that usually require human endeavor.
%	As expounded before, due to the special operating mechanism and capacity constraints of CAVs, potential safety hazards cannot be completely eliminated. 
%	There is still a need to deploy effective and reliable supervision approaches to ensure safe self-driving.
%	Meanwhile, as a result of the highly dynamic nature of networks, CAVs' driving status in multiple time slots are necessary to be combined and taken into account.
%	This yields time-related optimization and prediction problems. 
%	Accordingly, an incorporated assessment network with LSTM is proposed, as sketched in Fig.~\ref{EST}.
%	where $h_{t-1}$ and $h_{t}$ are the cell output at the previous and current moments, respectively. 
%	Analogously, $c_{t-1}$ and $c_{t}$ are the cell states at the previous and current moments, respectively.

\begin{figure*}[ht]
		\centering
		\includegraphics[width=1\textwidth]{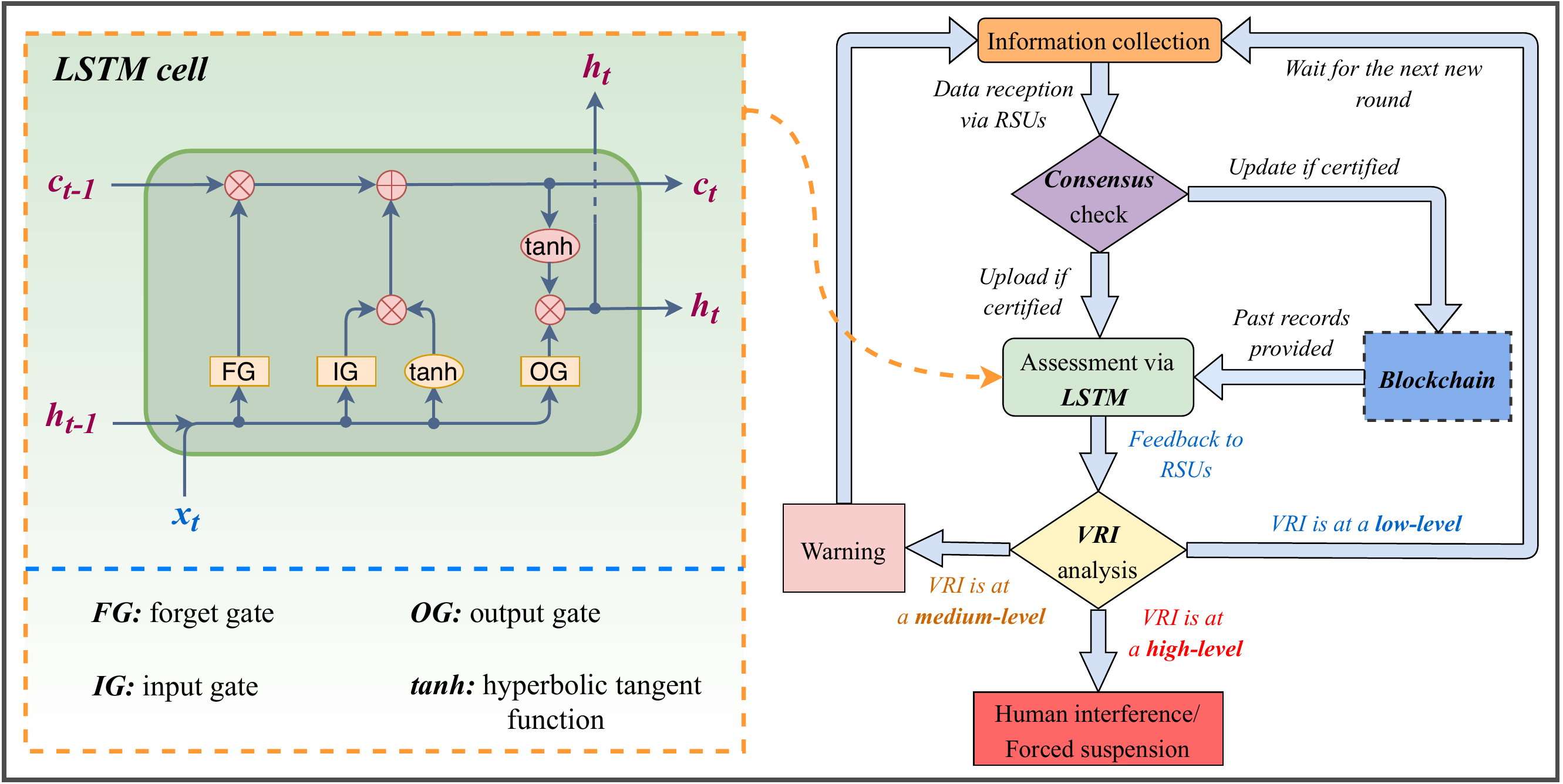} 
		\caption{The proposed EST scheme processing with LSTM network in the cloud.}
		\label{EST}
    \end{figure*}

	As expounded before, due to the special operating mechanism and capacity constraints of CAVs, potential safety hazards cannot be completely eliminated. There is still a need to deploy effective and reliable supervision approaches to ensure safe self-driving. Meanwhile, as a result of the highly dynamic nature of networks, CAVs’ driving status in multiple time slots are necessary to be combined and taken into account. This yields time-related optimization and prediction problems. Accordingly, an incorporated assessment network with LSTM is proposed, as sketched in Fig.~\ref{EST}.

	Generally, LSTM as an evolved gated recurrent neural network successfully overcomes the difficulties of long sequence time series dependence and gradient disappearance, where the technical details can be found in~\cite{sherstinsky2020fundamentals}. Hence, we exploit the LSTM to process time series vehicular data and extract useful information to complement our EST scheme. Meanwhile, multiple types of information exist in a moving CAV. Explicit features comprise, but are not limited to, its performed actions (e.g., velocity, accelerate, brake, turn), the safety level predicted at previous moments, and surrounding road conditions~\cite{singh2018branch}. Among them, RSUs are to gather the information recorded at current and past moments, before commencing the assessment process. Specifically, a performance metric, vehicle risk index (VRI), is defined to monitor different safety levels of CAV driving. It is worth mentioning that the definition of VRI is loosely coupled with the design of BEST framework. In other words, the definition and the way of calculating can only affect the absolute value of VRI, but does not invalidate the relative performance enhancement of BEST framework. Hence, any other sophisticated and accurate VRI models can be embedded in BEST. In the following, we demonstrate each phase in detail for a better understanding of EST in Fig.~\ref{EST}.
	
	\begin{itemize}[leftmargin=*]
		\item \emph{(Phase $1$) Information Sharing and Verification}: 
		First, each RSU acts as an information collector within its communication range to receive encrypted driving information (within a given time interval $T$) of all registered CAVs with their digital signatures $Sig_{V}$ and public keys $K^{u}_{V}$.
		Here, the $Sig_{V}$ and $K^{u}_{V}$ are used to verify the vehicle's identity. 
		Then, currently received data will be certified via a consensus protocol of blockchain to get authorization. Elaborations on this will be provided later.
		\item \emph{(Phase $2$) Dataset Preparation}:
		After authentications, RSUs will update their local database of blockchain and simultaneously read the past driving records of each vehicle to supplement datasets for LSTM.
		Then, RSUs upload prepared datasets to the cloud servers and wait for the assessment feedback.
		\item \emph{(Phase $3$) AI Assessment Process}:
		Exploiting cloud computing, time-series data-based regression problem can be rapidly solved by the fully trained LSTM model.
		Owing to our settings, the outcome of LSTM is $\left( VRI\| K^{u}_{V}\| Sig_{V}\right) $, where the VRI ($VRI\in \left( 0,1\right]$) indicates the hazardous degree of current driving.
		Afterward, this result will be fed back to the corresponding RSU of each CAV.
		\item \emph{(Phase $4$) VRI Analysis}:
		VRI represents the current safe driving circumstance of a moving CAV, where the lower the value, the safer the vehicle.
		Moreover, there should be different VRI thresholds considering the complex and dynamic road conditions in reality.
		Here, we take two standards as examples, i.e., a safe threshold of $\alpha$ and a dangerous threshold of $\beta$, respectively, where $0<\alpha <\beta <1$.
		\item \emph{(Phase $5$) Countermeasure Response}:
		According to the feedback, appropriate countermeasures are taken in time for the misbehaving CAVs with higher VRI values.
		For instance, an urgent warning should be given when VRI is at a medium level, i.e., $VRI\in \left(\alpha,\beta\right]$.
		Similarly, when $VRI\in \left(\beta,1\right]$, stricter measures (e.g., human intervention or forced suspension) should be executed immediately to prevent further serious consequences.
	\end{itemize}

	\begin{figure*}[ht]
		\centering
		\includegraphics[width=1\textwidth]{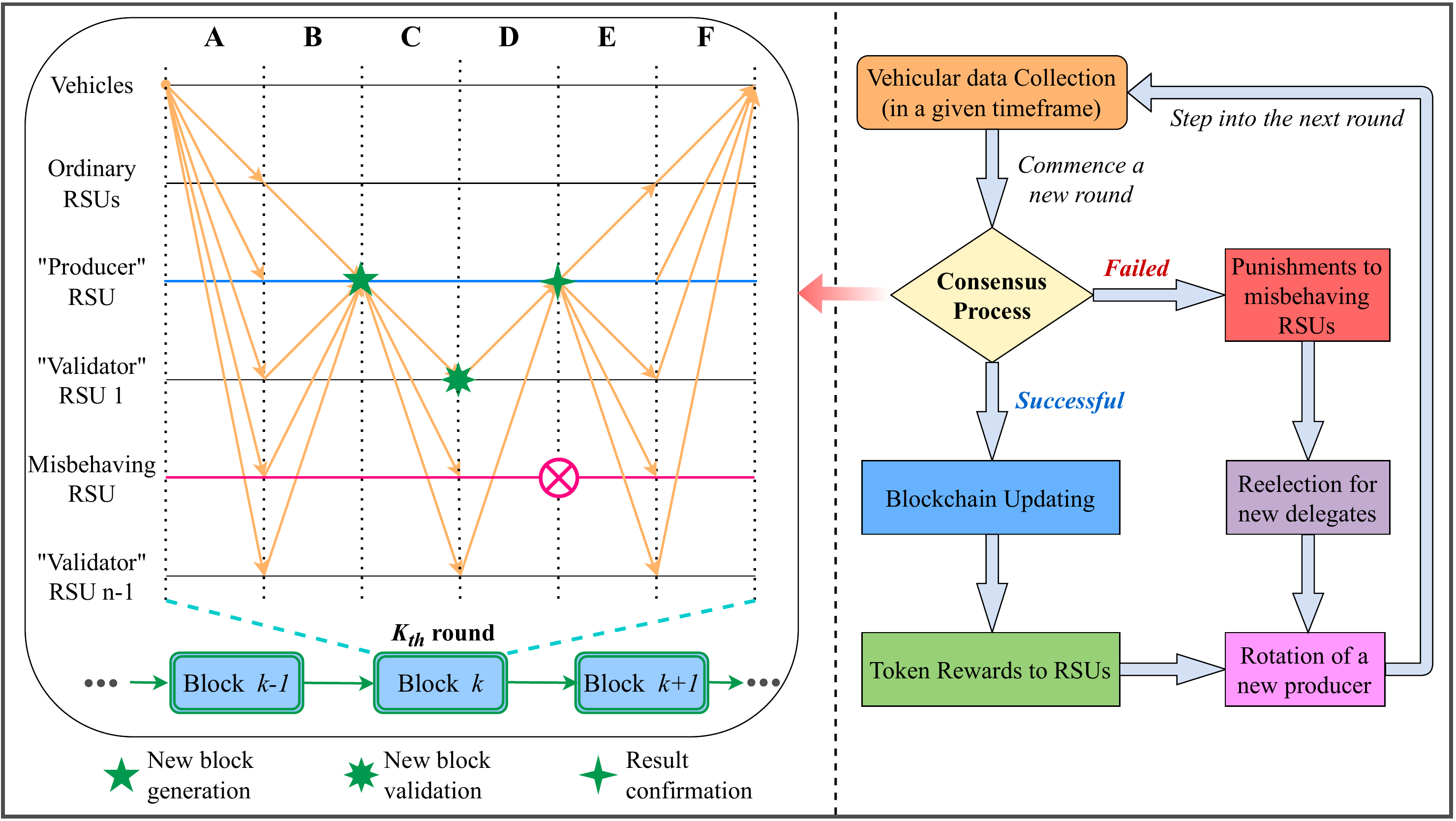} 
		\caption{The $K_{th}$ round of BFT-DPoS-based consensus process in a consortium blockchain: a) information sharing; b) records collection; c) new block broadcast; d) authentication feedback; e) blockchain update; f) data response.}
		\label{Blockchain}
    \end{figure*}

	\subsection{BFT-DPoS-Based Blockchain System}
%	Since the vehicular communication mainly relies on the cluster of approved RSUs that allow external clients to conduct data interaction in an authorized manner, a consortium blockchain is really suitable in this framework.
%%	First, choosing a fitting consensus mechanism is essential.
%	Generally, the most commonly used consensus mechanism in blockchain applications is Proof-of-Work (PoW) or Proof-of-Stake (PoS), however, neither of these two is the optimal alternative for autonomous driving.
%	PoW demands countless computation resources with considerable power consumption to complete mining tasks, which would create a burden on RSUs even with the help of MEC.
%	%In comparison, PoS determines the right of generating new blocks through calculating the percentage of tokens a participant hold and the duration of possession.
%	%Despite the increased speed in producing blocks, PoS still requires Hash calculation-based mining operation with global validation, resulting in weak supervision and low efficiency.
%	Besides, PoS requires Hash calculation-based mining operation with global validation, resulting in weak supervision and low efficiency.
%	Besides, a centralization phenomenon may further emerge in extreme cases owing to PoS's stake holding mechanism.
	Since the vehicular communication mainly relies on the cluster of approved RSUs that allow external clients to conduct data interaction in an authorized manner, a consortium blockchain is really suitable in this framework. Generally, the most commonly used consensus mechanism in blockchain applications is Proof-of-Work (PoW) or Proof-of-Stake (PoS).
	However, neither of these is the optimal alternative for autonomous driving. PoW demands countless computation resources with considerable power consumption to complete mining tasks, which would create a burden on RSUs, even with the help of MEC. Besides, PoS requires hash calculation-based mining operation with global validation, resulting in weak supervision and low efficiency.
	
%	For a better collaboration with EST, a BFT-DPoS consensus protocol is applied in our blockchain, as elucidated in Fig.~\ref{Blockchain}, which ensures excellent transaction throughput necessary to support real-time operations in the CAV network.
%	As an exemplification, cryptocurrency EOS leverages BFT-DPoS to reach an irreversible consensus within only $1$ second~\cite{io2018eos}.
%	Specifically, DPoS is a democratic form of PoS based on the consensus nodes (i.e., CRNs group) voted via public delegation (i.e., all RSUs).
%%	In other words, each RSU can be regarded as a token holder to cast a vote and entrust its own stake to a delegate as its proxy.
%	Once finalizing a round of delegation procedure, CRNs are able to exercise their authorities of ledger management.
%	Moreover, by incorporating an extra layer of BFT, DPoS mechanism can further guarantee an ultra-robust and highly-valid blockchain with low consensus delay~\cite{xiao2020survey}.
%	To elaborate further, we give the workflow of BFT-DPoS process as follows:
	For a better collaboration with EST, a BFT-DPoS consensus protocol is applied in our blockchain, as shown in Fig. 4, which ensures excellent transaction throughput necessary to support real-time operations in the CAV network. As an exemplification, cryptocurrency EOS leverages BFT-DPoS to reach an irreversible consensus within only 1 second~\cite{io2018eos}. Specifically, DPoS is a democratic form of PoS based on the consensus nodes (i.e., CRNs group) voted via public delegation (i.e., all RSUs). Once finalizing a round of delegation procedure, CRNs are able to exercise their authorities of ledger management. Moreover, by incorporating an extra layer of BFT, DPoS mechanism can further guarantee a robust and highly valid blockchain with low consensus delay~\cite{xiao2020survey}. To elaborate further, we give the workflow of BFT-DPoS process as follows.
	
	\begin{itemize}[leftmargin=*]
		\item \emph{Preparations}:
		Initially, the network elects several most trusted RSUs as CRNs based on the token deposits proportion voted in a stake pool, where more details can be discovered in~\cite{xiao2020survey}. The rest ones become the ordinary nodes (i.e., ORNs). They are only responsible for data interaction and blockchain storage. Next, a new round of consensus process is capable for commencement.
%		Initially, the network elects several most trusted RSUs as CRNs based on the token deposits proportion voted in a stake pool, where more details can be discovered in~\cite{xiao2020survey}.
%		The rest ones become the ordinary nodes (i.e., ORNs) who are only responsible for data interaction and blockchain storage.
%		Next, a new round of consensus process is capable for commencement.
		
		\item \emph{(Step $1$) Block Producer Election}:
		According to the stake information fetched from all CRNs, a pseudorandom sequence of block generation opportunities is first generated. Correspondingly, each CRN is elected as a~\emph{producer} to propose new blocks in a round-robin fashion, while the others act as \emph{validators} for auditing the new block at the same time.
		
		\item \emph{(Step $2$) New Block Generation}:
		The producer collects all records of vehicular driving status that occurred within $T$, then uses its private key to encrypt and pack them into a new block.
		Meanwhile, producer's signature $Sig_{pro}$ with its public key is also attached to insure that validators can confirm the block source.
		
		\item \emph{(Step $3$) New Block Validation}:
		The BFT-DPoS enables the producer to broadcast new block to all validators at once, which replaces the traditional approach of sequential validation in DPoS and significantly promotes the validation efficiency.
		After that, each validator compares the received duplicated block with local replicas to verify the authenticity and feed the result with its signature $Sig_{va}$ back to producer.
		
		\item \emph{(Step $4$) Result Confirmation}:
		Based on the BFT rule, when exceeding $2/3$ different signed blocks are received by the producer~\cite{xiao2020survey}, this new block is deemed valid and irreversible.
		Otherwise, the system will forcibly suspend the current procedure and return to the \emph{Step $1$} to prepare for the next new round of consensus. 
		
		\item \emph{(Step $5$) Blockchain Extension}:
		After confirming that the new block is valid, the producer conducts the second broadcast to RSUs (both CRNs and ORNs) to complete the blockchain update.
		In the meantime, a new round of consensus process will commence from the next producer in the established sequence.
		Consequently, the driving records gathered by RSUs can be uploaded to the EST with authorization.
		
		\item \emph{Rewards and Punishments}:
		To enforce integrity and credibility in blockchain, a reward and punishment-based incentive mechanism is devised to encourage trustworthy delegation and consensus participation.
		After each round, CRNs receive token rewards proportional to the deposits they voted. 
		This rule is also applied to the ORNs to gain some dividends.
		However, the CRNs with misbehaviors will be confronted with the risks of voting out and token deduction.
		If one RSU is removed from the committee, a new replacement will be reelected from the ORNs to fill the vacancy.
	\end{itemize}
	
%	In summary, the proposed BEST can offer sufficient security protection for vehicular data and driving.
%	Nevertheless, since the delay and communication overhead cannot be neglected in such a time-critical driving scenario, we briefly analyze their impact in the BEST here for clearer understanding.
%	First, the communication delay in BEST is the same when compared to conventional CAV networks, because no further burden here is imposed on the communication interplay phase.
%	For the computing delay, note that no extra delay is caused by the EST scheme, thanks to the pre-training mechanism that can make a well-trained LSTM be directly used for prediction.
%	Meanwhile, we also deploy powerful MEC serves, as aforementioned, to offload most blockchain tasks from the RSUs and greatly reduce the computing delay.
%	Besides, the BFT-DPoS consensus we choose is a very efficient mechanism ($0.5s$ per block~\cite{io2018eos}), where the delay can be considered tolerable in the BEST scenario at such a fast block generation speed.
%	As for the communication overhead, it can be observed in Fig.~\ref{Blockchain} that only a small amount of signaling interaction is demanded for blockchain to finalize a round of consensus confirmation, while only several bits are required for each signaling transmission.
%	In this context, either the delay or the communication overhead issue can be well tackled in a BEST-enabled CAV network.
	In summary, the proposed BEST can offer sufficient security protection for vehicular data and driving. Nevertheless, since the delay and communication overhead cannot be neglected in such a time-critical driving scenario, we briefly analyze their impact in the BEST for clearer understanding. First, the communication delay in BEST is the same when compared to conventional CAV networks because no further burden is imposed on the communication interplay phase. For the computing delay, note that no extra delay is caused by the EST scheme, thanks to the pre-training mechanism that can make a well-trained LSTM be directly used for prediction. Meanwhile, we also deploy powerful MEC serves, as aforementioned, to offload most blockchain tasks from the RSUs and greatly reduce the computing delay. Besides, the BFT-DPoS consensus we choose is a very efficient mechanism ($0.5$ seconds per block~\cite{io2018eos}), where the delay can be considered tolerable in the BEST scenario at such a fast block generation speed. As for the communication overhead, it can be observed in Fig.~\ref{Blockchain}, where only a small amount of signaling interaction is demanded for blockchain to finalize a round of consensus confirmation, and while only several bits are required for each signaling transmission. In this context, either the delay or the communication overhead issue can be well tackled in a BEST-enabled CAV network.

	\section{Simulation Results and Analysis}
	In this section, simulations are presented to demonstrate the performance of the proposed BEST (i.e., blockchain with LSTM) when compared with three other combination schemes between~\textit{LSTM, deep neural network (DNN), blockchain, and centralization approaches}, in environments involving multiple malicious CAVs.
	Herein, the centralization approach indicates that all data from CAVs are managed solely by a central entity, which is only accountable for storing the uploaded information, and adopting the conventional data encryption method like cryptography but without any participation of blockchain.
	Besides, we choose the most commonly used DNN model as the benchmark of LSTM, where the aforementioned VRI metric is set as the target monitoring parameter.
	
	In our simulations, the BEST framework is implemented in a computer with six CPU cores and Inter Core i7 processor, while the main software environment is Tensorflow $2.1.0$ and Python $3.7$.
	We first simulate a general CAV network scenario, in which the numbers of RSUs and CAVs are set to $20$ and $300$, respectively.
	Specifically, RSUs are set in a given area and have fixed positions, while the CAVs randomly locate and each CAV is associated with their nearest RSU for simplicity.
	For each CAV, the initial velocity is randomly distributed between $0\sim 50 km/h$ with a steady acceleration between $-10\sim 10 m/s^{2}$.
	Note that these parameters are only as the input status data without affecting any performance of the BEST.
	Some other status information, including the amount of neighboring vehicles, minimum distance to other vehicles, and position, etc., is also shared with its adjacent RSU per second.
	Herein, we assume that all CAVs have fixed driving directions, and all driving status data within $10$ seconds of the CAVs are then collected for subsequent analysis.
	For VRI calculation, we initially set a total of four risk levels (low, medium, high, and accident level) for CAVs in the simulation, thereby becoming a classification optimization problem (i.e., cross-entropy is adopted as the loss function).
	For blockchain system, the RSU-enabled consensus rule is considered and deployed in this network, where the block size is $8$ MB and the maximum block interval is set as $1$ second. 
	Meanwhile, a two-layer LSTM is constructed to predict VRI with a comparison object of a four-layer DNN.
	Dropout technique is also exploited to avoid the overfitting problem and an Adam optimizer is utilized for gradient updates.
	
	\begin{figure}[t]
		\centering
		\includegraphics[width=0.5\textwidth]{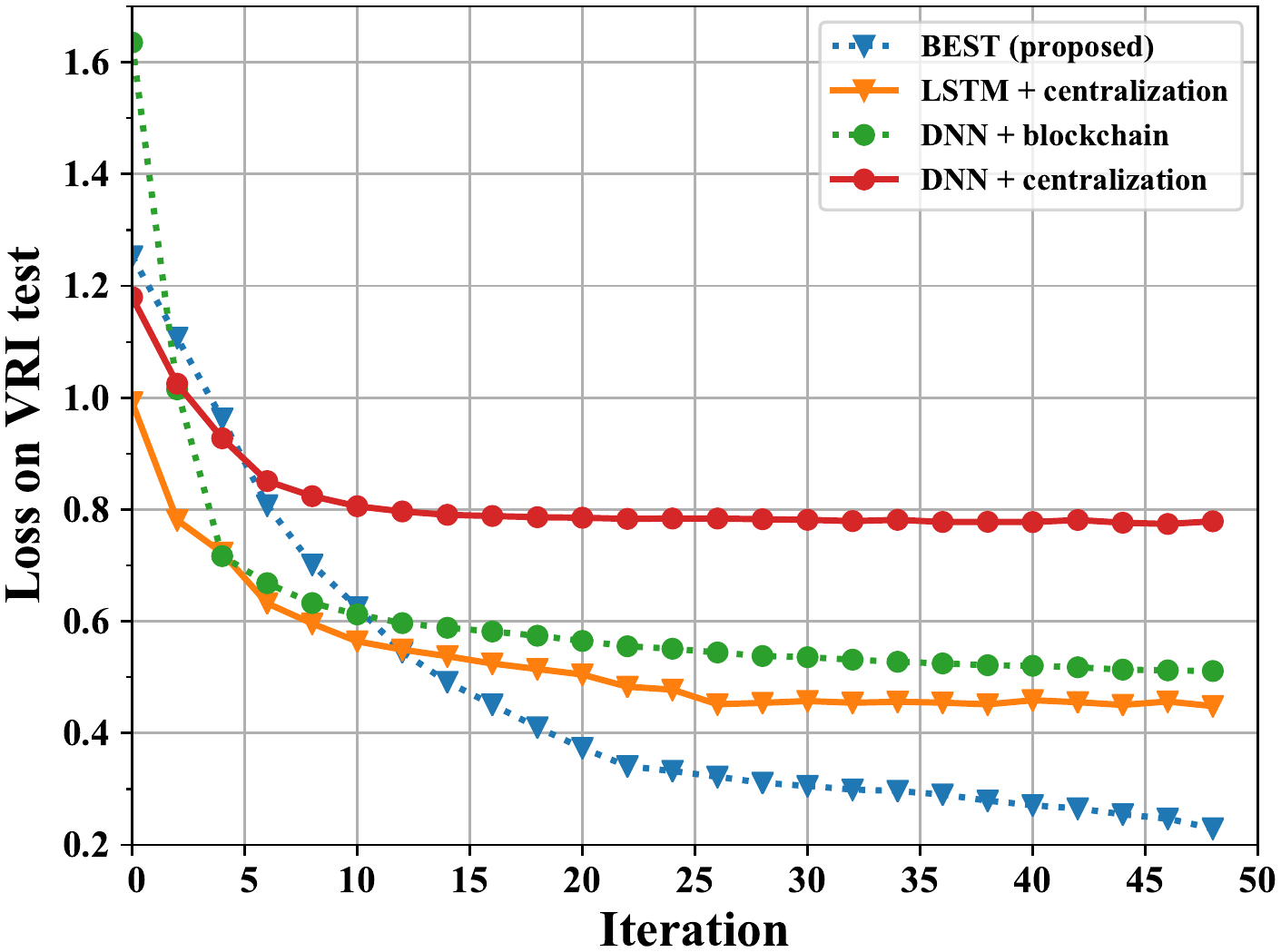} 
		\caption{VRI prediction loss \emph{vs.} training iteration with different approaches.}
		\label{Simulation1}
    \end{figure}
    
    \begin{figure}[t]
		\centering
		\includegraphics[width=0.5\textwidth]{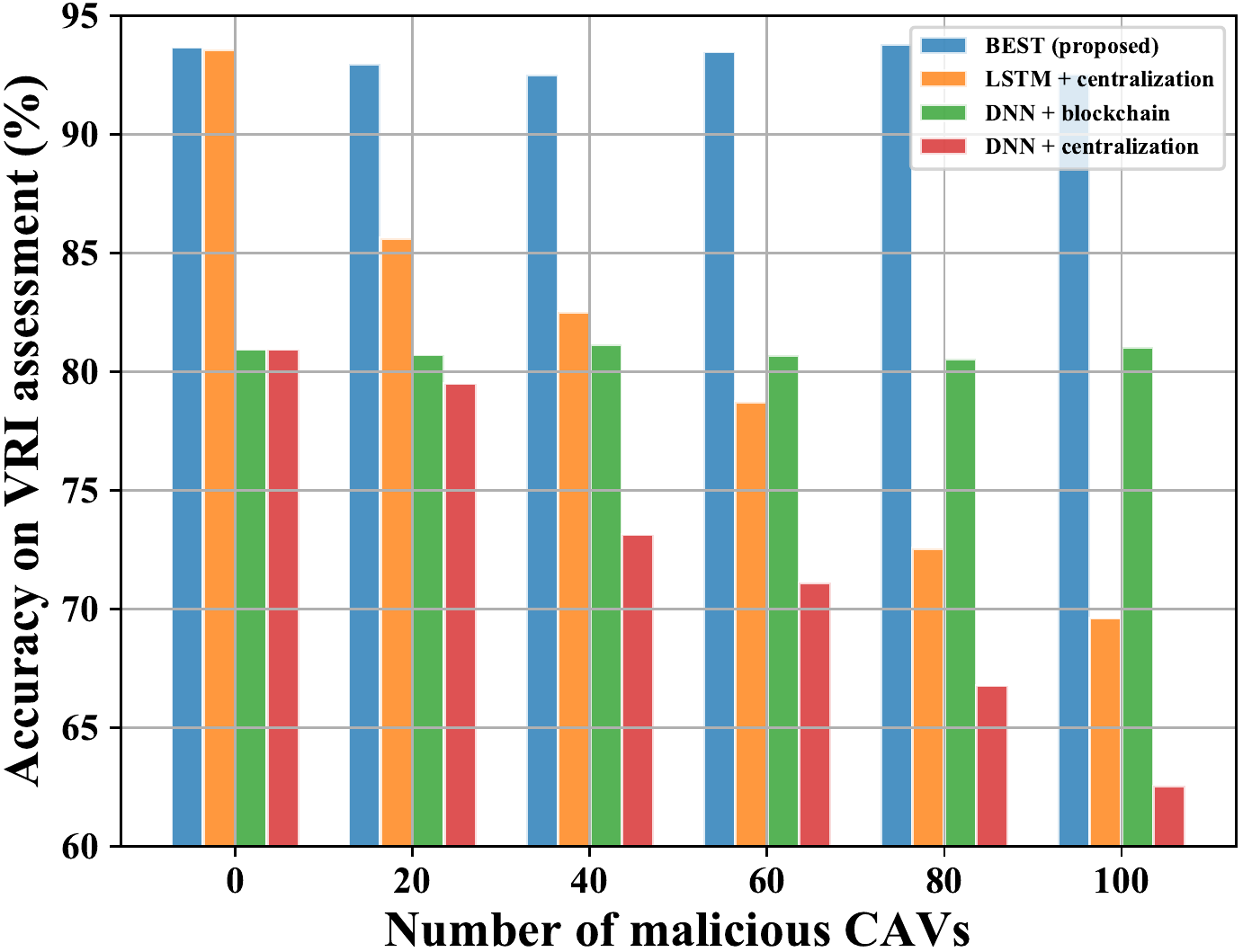} 
		\caption{Accuracy comparison of VRI assessment under different numbers of malicious CAVs.}
		\label{Simulation2}
    \end{figure}
	
	As depicted in Fig.~\ref{Simulation1}, we first show the loss convergence results for VRI predictions under the four different schemes.
	Obviously, the proposed integrated framework of LSTM with blockchain achieves the optimal VRI prediction loss of around $0.2$ in $50$ training iterations, which far outperforms the case using DNN instead. 
	Notably, we add $50$ malicious CAVs in this figure to continuously forge fake, meaningless, and dishonest content, and upload them to RSUs for causing chaos to the training dataset.
	This setting aims to better test and compare their respective tolerance to malicious content attacks.
	By observing the curve trends of centralized approaches, it seems that traditional cryptography cannot effectively preclude the negative impacts from fake content attacks, causing a high prediction loss.
	Nevertheless, the schemes using the blockchain technique can significantly identify and eliminate false content in the contaminated dataset and attain very low loss with fast convergence speed.
	
	Next, we further compare the accuracy of VRI assessment with different numbers of malicious CAVs participating in the network as shown in Fig.~\ref{Simulation2}, thus to investigate the robustness of the four schemes. 
%	we further investigate the robustness divergence between these schemes on VRI assessment , when there are different numbers of malicious CAVs participating in the network.
	It can be clearly seen that the proposed BEST always maintains high prediction accuracy of $92.5\sim 93.8\%$, while the accuracies of other approaches are far below it (all lower than $85.6\%$).
	For instance, the VRI assessment performance of centralized approaches drops drastically as the number of malicious CAVs increases from $0$ to $100$.
	Due to its inability to authenticate fake data, the performance apparently becomes poor when many malicious CAVs exist.
	In contrast, the blockchain method can successfully detect and eliminate these fake data via its powerful consensus rule and unforgeable unified ledger, greatly improving the purity of dataset and ensuring a relatively stable precision level for AI prediction.
	This conclusion is also highly consistent with the results obtained in Fig.~\ref{Simulation1}.
	In summary, instantiations that rely on our BEST framework can smoothly realize safe-driving assessment at a very high accuracy, while gaining adequate data credibility and security in autonomous driving networks.
	
	\section{Open Challenges and Discussions}
%	In spite of many superiorities, the proposed BEST framework still imposes some associated and nontrivial challenges that should be discussed before unlocking its full potentials.
	In spite of many superiorities, the proposed BEST framework still imposes some associated and nontrivial challenges that should be discussed before unlocking its full potential.
	
	\textbf{Inactive Information Sharing:}
	Since the a al effect of the proposed framework primarily depends on the information shared by CAVs in the communication community, vehicles may lack the enthusiasm to upload their data to RSUs without ample compensation. Therefore, a rewards-based incentive mechanism for CAVs can be embedded into the BEST to encourage vehicles to spontaneously share information and attract other vehicles to participate in this framework.
%	Since the actual effect of the proposed framework primarily depends on the information shared by CAVs in the communication community, vehicles may lack the enthusiasm to upload their data to RSUs without ample compensation.
%	Therefore, a rewards-based incentive mechanism for CAVs can be embedded into the BEST to encourage vehicles to spontaneously share information and attract more other vehicles to participate in this framework.
	
	\textbf{Highly Dynamic Road Conditions:}
	The road conditions of different RSUs vary according to their locations in the city; traffic congestion under the same RSU in different time periods is also distinct. This fact leads to an imbalance of task allocation in BEST, where an excessive volume of vehicular contents may be sent to a single RSU while other RSUs only receive a few. To this end, proposing a real-time task scheduling scheme for RSUs can promote the effectiveness of information gathering for BEST under high road condition dynamics.
%	The road conditions of different RSUs vary according to their locations in the city, and traffic congestion under the same RSU in different time periods is also distinct.
%	This fact leads to an imbalance of task allocation in BEST, where an excessive volume of vehicular contents may be sent to a single RSU while some other RSUs only receive a few. 
%	To this end, proposing a real-time task scheduling scheme for RSUs can promote the effectiveness of information gathering for BEST under high road condition dynamics.
	
	\textbf{Resources Allocation in CAVs:}
%	Connecting to RSUs or other vehicles is essential for a CAV to share data to BEST with resources allocation. 
%	It is inevitable to consume a certain amount of wireless communication resources for data collection, signal transmission, and information sharing in BEST.
%	However, due to the limited resources, CAV networks have to well allocate them across multiple devices with different communications tasks to reach an optimal resource utilization.
%	In this case, reinforcement learning algorithm might be a promising solution to automatically and smartly achieves resource allocation for each operation in moving CAVs.
	It is inevitable to consume a certain amount of wireless communication resources for data collection, signal transmission, and information sharing in BEST. However, due to the limited resources, CAV networks have to allocate them across multiple devices with different communication tasks to reach an optimal resource utilization. In this case, reinforcement learning algorithms might be a promising solution to automatically and smartly achieve resource allocation for each operation in moving CAVs.
	
	\section{Conclusions}
%	In this article, we proposed a novel BEST framework that incorporates AI and consortium blockchain, offering driving safety and data security simultaneously in CAV networks.
%	An LSTM model was applied in the EST scheme for evaluating the VRI, followed by a blockchain system for a supplement of data credibility through its powerful consensus mechanism.
%	Simulation results further demonstrated that our BEST could maintain significantly high accuracy of driving risk assessment when compared to existing schemes, even if there is a high amount of false content interference from malicious CAVs.
%	Finally, several open challenges and potential solutions were discussed.
%	We hope that this work becomes a pioneer in building an efficient and reliable supervision system based on AI and blockchain to underpin future autonomous driving applications.
	In this article, we proposed a novel BEST framework that incorporates AI and consortium blockchain, offering driving safety and data security simultaneously in CAV networks. An LSTM model was applied in the EST scheme for evaluating the VRI, followed by a blockchain system for a supplement of data credibility through its powerful consensus mechanism. Simulation results further demonstrated that our BEST could maintain significantly high accuracy of driving risk assessment when compared to existing schemes, even if there is a large amount of false content interference from malicious CAVs. Finally, several open challenges and potential solutions were discussed. We hope that this work becomes a pioneer in building an efficient and reliable supervision system based on AI and blockchain to underpin future autonomous driving applications.
	
	\section{Acknowledgements}
	This work was supported by the PETRAS National Centre of Excellence for IoT Systems Cybersecurity, which has been funded by the UK EPSRC under grant number EP/S035362/1. We would like to recognize the studentship support from AIDrivers Ltd.
	
	\bibliographystyle{IEEEtran}
	\bibliography{main}

\begin{thebibliography}{10}

\bibitem{abboud2016interworking}
K. Abboud, H.~A. Omar, and W. Zhuang, ``Interworking of DSRC and cellular network technologies for V2X communications: A survey,''
\newblock {\em IEEE transactions on vehicular technology}, vol.~65, no.~12, pp.~9457--9470, 2016.

\bibitem{de2020driverless}
G. De La Torre, P. Rad, and K.-K.~R. Choo, ``Driverless vehicle security: Challenges and future research opportunities,''
\newblock {\em Future Generation Computer Systems}, vol.~108, pp.~1092--1111, 2020.

\bibitem{wang2018networking}
J. Wang, J. Liu, and N. Kato, ``Networking and communications in autonomous driving: A survey,''
\newblock {\em IEEE Communications Surveys \& Tutorials}, vol.~21, no.~2, pp.~1243--1274,
  2018.

\bibitem{sae2014taxonomy}
{SAE On-Road Automated Vehicle Standards Committee}, ``Taxonomy and definitions for terms related to on-road motor vehicle automated driving systems,''
\newblock {\em SAE Standard J}, vol.~3016, pp.~1--16, 2014.

\bibitem{efrati2018uber}
A. Efrati,
``Uber finds deadly accident likely caused by software set to ignore
  objects on road,''
\newblock {\em The information}, 2018.

\bibitem{xing2019trust}
R. Xing, Z. Su, N. Zhang, Y. Peng, H. Pu, and J. Luo,
``Trust-evaluation-based intrusion detection and reinforcement learning
  in autonomous driving,''
\newblock {\em IEEE Network}, vol.~33, no.~5, pp.~54--60, 2019.

\bibitem{ye2021joint}
Q. Ye, W. Shi, K. Qu, H. He, W. Zhuang, and X.
  Shen,
``Joint RAN slicing and computation offloading for autonomous vehicular networks: A learning-assisted hierarchical approach,''
\newblock {\em IEEE Open Journal of Vehicular Technology}, vol.~2, pp.~272--288, 2021.

\bibitem{posner2021federated}
J. Posner, L. Tseng, M. Aloqaily, and Y.
  Jararweh,
``Federated learning in vehicular networks: opportunities and solutions,''
\newblock {\em IEEE Network}, vol.~35, no.~2, pp.~152--159, 2021.

\bibitem{cui2019review}
J. Cui, L.~S. Liew, G. Sabaliauskaite, and F. Zhou,
``A review on safety failures, security attacks, and available
  countermeasures for autonomous vehicles,''
\newblock {\em Ad Hoc Networks}, vol.~90, p.~101823, 2019.

\bibitem{jie2020game}
Y. Jie, C.~Z. Liu, M. Li, K.-K.~R. Choo, L. Chen, and C. Guo,
``Game theoretic resource allocation model for designing effective traffic safety solution against drunk driving,''
\newblock {\em Applied Mathematics and Computation}, vol.~376, p.~125142,
  2020.

\bibitem{zargar2013survey}
S.~T. Zargar, J. Joshi, and D. Tipper,
``A survey of defense mechanisms against distributed denial of service (DDoS) flooding attacks,''
\newblock {\em IEEE communications surveys \& tutorials}, vol.~15, no.~4, pp.~2046--2069,
  2013.

\bibitem{sherstinsky2020fundamentals}
A. Sherstinsky,
``Fundamentals of recurrent neural network (RNN) and long short-term
  memory (LSTM) network,''
\newblock {\em Physica D: Nonlinear Phenomena}, vol.~404, p.~132306, 2020.

\bibitem{singh2018branch}
M. Singh and S. Kim,
``Branch based blockchain technology in intelligent vehicle,''
\newblock {\em Computer Networks}, vol.~145, pp.~219--231, 2018.

\bibitem{io2018eos}
{IO, EOS},
``EOS. IO technical white paper v2,''
\newblock {\em EOS, Tech. Rep., March}, 2018.

\bibitem{xiao2020survey}
Y. Xiao, N. Zhang, W. Lou, and Y.~T. Hou,
``A survey of distributed consensus protocols for blockchain networks,''
\newblock {\em IEEE Communications Surveys \& Tutorials}, vol.~22, no.~2, pp.~1432--1465,
  2020.

\end{thebibliography}
	\vspace{-35pt}
	\begin{IEEEbiographynophoto}{Le Xia} (l.xia.2@research.gla.ac.uk)
	 received his B.Eng. degree and M.Eng. degree in electronics and communication engineering from the University of Electronic Science and Technology of China (UESTC) in 2017 and 2020, respectively. He is pursuing his Ph.D in the James Watt School of Engineering, University of Glasgow, United Kingdom. His research interests include driverless vehicular networks and intelligent wireless communications.
	\end{IEEEbiographynophoto}
	\vspace{-35pt}
	\begin{IEEEbiographynophoto}{Yao Sun} (Yao.Sun@glasgow.ac.uk)
	 is currently a Lecturer with James Watt School of Engineering, the University of Glasgow, UK. He has won the IEEE Communication Society of TAOS Best Paper Award in 2019 ICC. His research interests include intelligent wireless networking, blockchain system, and resource management in mobile networks.
	\end{IEEEbiographynophoto}
	\vspace{-35pt}
	\begin{IEEEbiographynophoto}{Rafiq Swash} (Rafiq.Swash@brunel.ac.uk)
	 is the founder of AIDrivers Ltd., a lecturer with Brunel University London, and also a visiting professor with Changchun Institute of Optics. He has given scientific talks in number of international scientific and innovation conferences as a keynote speaker in Europe, China, Qatar, India, and UAE.
	\end{IEEEbiographynophoto}
	\vspace{-35pt}
	\begin{IEEEbiographynophoto}{Lina Mohjazi} (Lina.Mohjazi@glasgow.ac.uk)
	 is a Lecturer in the James Watt School of Engineering, University of Glasgow, UK. She received her Ph.D. degree from the University of Surrey, UK, in 2018. Her research interests include beyond 5G wireless technologies, wireless power transfer, machine learning, and reconfigurable intelligent surfaces.
	\end{IEEEbiographynophoto}
	\vspace{-35pt}
	\begin{IEEEbiographynophoto}{Lei Zhang} (Lei.Zhang@glasgow.ac.uk)
	is a Senior Lecturer at the University of Glasgow, U.K. His research interests include wireless communication systems and networks, blockchain technology, data privacy and security, etc. He is an associate editor of IEEE Internet of Things (IoT) Journal, IEEE Wireless Communications Letters, and Digital Communications and Networks.
	\end{IEEEbiographynophoto}
	\vspace{-35pt}
	\begin{IEEEbiographynophoto}{Muhammad Ali Imran} (Muhammad.Imran@glasgow.ac.uk) 
	is a Professor of communication systems with the University of Glasgow, UK, and a Dean with Glasgow College UESTC. He is also an Affiliate Professor with the University of Oklahoma, USA, and a Visiting Professor at University of Surrey, UK. He has over 20 years of combined academic and industry experience with several leading roles in multi-million pounds funded projects.	\end{IEEEbiographynophoto}

\end{document}